%% file: percon14.tex
\documentclass[a4paper]{sigchi}

\usepackage{authblk}
\makeatletter
\renewcommand\AB@affilsepx{\protect\Affilfont}
\makeatother

\usepackage[hyphens]{url}
\usepackage[hidelinks]{hyperref}

\usepackage{graphicx}
\usepackage{epstopdf}

\usepackage{color}

\newcommand{\one}{({\em i})}
\newcommand{\two}{({\em ii})}

\usepackage{xspace}
\newcommand{\databox}{Databox\xspace}

\usepackage[pass]{geometry}
\def\etal{{\it et al.}}
\date{}

\usepackage{enumitem}
\setitemize{noitemsep,topsep=0pt,parsep=0pt,partopsep=0pt,leftmargin=*}
\setenumerate{noitemsep,topsep=0pt,parsep=0pt,partopsep=0pt,leftmargin=*}

\begin{document}
\title{Personal Data: Thinking Inside the Box}
\toappear{}


\author[1]{Hamed Haddadi}
\author[2]{Heidi Howard}
\author[2]{Amir Chaudhry}
\author[2]{Jon Crowcroft}
\author[2]{Anil Madhavapeddy}
\author[2]{Richard Mortier}

\affil[1]{Queen Mary University of London }
\affil[2]{ Computer Laboratory, University of Cambridge}

\maketitle

\input{intro}
\input{motivation}
\input{databox}
\input{contents}
\input{where}
\input{conclusion}

\paragraph{Acknowledgements}
We appreciate constructive feedback on this paper from Ian Brown (University of Oxford), Essam Mansour (QCRI), Irene Ng (University of Warwick), Paul Francis (MPI-SWS), and Jeremy Yallop (University of Cambridge).

\footnotesize
{
  \bibliographystyle{acm-sigchi}
  \bibliography{percon14}
  \balancecolumns
}
\end{document}

%% file: intro.tex
\section{Introduction}
\label{s:intro}


We are in the middle of a `personal data gold rush' driven by the dominance of advertising as the primary source of revenue for most online companies. Internet services, advertisers, and even governments are all casting a wide net to accumulate personal data about individuals. This accumulation is generally occurring with minimal consideration of \emph{us}, the individuals at the heart of this process.

Governments and regulatory bodies, such as the European Union, have attempted to impose regulatory frameworks that force the market to recognise certain rights of individuals. Unfortunately, legal systems are not sufficiently agile to keep up with the rapid pace of change in this area. Self-regulation proposals such as the \emph{Do Not Track} headers\footnote{\url{http://donottrack.us/}} have been ineffective in reducing behavioural targeting and advertising. For example, in a 2012 study by Balebako \etal~\cite{Balebako12measuringthe}, only two of the thousands of existing advertising agencies had agreed to respect the headers. This number has since grown to 20,\footnote{\url{http://donottrack.us/implementations}} but this remains an insignificant fraction of such services~\cite{DBLP:journals/corr/FalahrastegarHUM14}. Fundamentally, imposing constraints that ignore the interests of advertisers and analytics providers, in many cases the business models that drive ``free'' web services and mobile apps, is likely to fail~\cite{dontkill, narseoIMC}. This further reinforces the notion that \emph{if you are not paying for it, you are the product}.\footnote{Although nothing precludes being both a paying customer \emph{and} the product being sold.}

A range of personal data technology startups have been formed in recent years, in response to growing public awareness about how our data is processed. These aim to put users explicitly in control of their personal data (or metadata, see openPDS~\cite{10.1371/journal.pone.0098790}), providing platforms through which they can permit advertisers and content providers to enjoy metered access to valuable personal data. In exchange, participating users could potentially benefit by receiving a portion of the monetary value generated from their data as it is traded in an increasingly complex ecosystem~\cite{DBLP:journals/corr/FalahrastegarHUM14}. Unfortunately, all these approaches provide both a logical \emph{and a physical} single point of control over our personal data: typically, they entail lodging information in the cloud where the service is running. This naturally leads to a host of trust issues for users, who find themselves not just having to trust the service directly but also the infrastructure providers involved, other parties such as local law enforcement in the specific jurisdiction, and the possibility of collusion between these cloud services to build ever more detailed models of individuals.


Individuals' responses to this are largely complex and context dependent~\cite{dourish-context}: for example, Westin classifies people by whether they are \emph{privacy unconcerned}, \emph{privacy fundamentalists}, or \emph{privacy pragmatists}~\cite{westin-survey}. He finds that 16\% are unconcerned, 24\% are fundamentalist, and 60\% of all respondents fall into the last category, where attitudes to privacy are dependent on a wide range of specifics such as the particular data and its relevance, the industry involved, and so on. Though this form of classification has been recently challenged~\cite{Vulnerable2014}, in essence it demonstrates that `personal data' is inherently social: it is generally not a practical response to decide to withdraw completely from all online activity.

We propose there is a need for a technical platform enabling people to engage with the collection, management and consumption of personal data; and that this platform should itself be personal, under the direct control of the individual whose data it holds.\footnote{We accept that the notion of individual ownership of a fundamentally shared asset such as data is itself problematic, and we will discuss this in more detail subsequently.} In what follows, we refer to this platform as the \databox, a \emph{personal}, \emph{networked service} that collates personal data and can be used to make those data available. While your \databox is likely to be a \emph{virtual} platform, in that it will involve multiple devices and services, at least one instance of it will exist in physical form such as on a physical form-factor computing device with associated storage and networking, such as a home hub.

%% file: motivation.tex
\section{Why do we need a Databox?}
\label{s:motivation}


Today, many businesses rely on personal data while many services (notably online social networks) operate as \emph{walled gardens} --- increasing \emph{lock-in} and network externalities are preventing formation of a truly competitive market. In addition, regular data leakages and the privacy issues of cloud-based data silos (see recent media reports concerning account data and password leakage by Dropbox and Skype), the opaque nature of data inferences constructed by advertising agencies, and the trade of cookies and personal data between third parties, all call for means to index and control our ever-increasing portfolio of personal data.

The sheer amount of activity in this sector suggests that there is at least \emph{some} unmet need here, though it is perhaps less clear precisely what that is. In particular, why would we \emph{each} want a \databox. One very practical motivation is the range of privacy threats that arise due to, for example, the range and reach of the information being stored about us by third-party websites~\cite{DBLP:journals/corr/FalahrastegarHUM14}.\footnote{That is, websites that we do not interact with explicitly but which are invoked by `first party` websites we actually visit.} This is in addition to privacy threats from data aggregators over which users have no control, including government agencies interested in surveillance such as the NSA and GCHQ, advertisers and credit scoring companies.


More importantly, one of the benefits of having a \databox system would be to enable a decentralised platform that other developers can target to provide services and software on. In the current world of centralised silos, all Independent Software Vendors (ISV) are at the whim of the large platform and API providers. Users and ISVs suffer whenever there is a conflict of interest with those providers, which hampers innovation and can distort markets. Platforms like the \databox will not replace dedicated, application-specific services such as Facebook and Gmail, and neither are they oriented solely towards \emph{privacy}, and the prevention of activities involving personal data. Rather, they enable \emph{new} applications able to combine data from many silos to draw inferences unavailable in the existing marketplace. At the same time, they provide for the HDI concepts of legibility, agency and negotiability, going some way to redress the highly asymmetric power relationship that pertains currently in the personal data ecosystem. This potentially opens up a range of market and social approaches to the ways in which we conceive of, manage and exploit ``our'' data.

A host of other motivations and uses for such a \databox have been presented elsewhere~\cite{Mun:2010:PDV:1921168.1921191,mortier2010personal,guha.hotnets09}. These include privacy-preserving advertising, market research, health applications, \textit{Quantified Self}, and personal archives. Indeed, an alternative, or additional, regulatory response in form of interventions could lead to increased competition in the data market. In aggregate, these examples point to a need for individuals to have tools that allow them to take more explicit control over the collection and consumption of their data and the information inferred from their online activities. However, given the reliance of the existing web ecosystem on the use of personal data, it is important that evolutionary paths are available to enable the ecosystem to survive. For example, control mechanisms may include the ability for individuals to sell or donate their data, in whole or in part~\cite{privalytics}.

Mapping these motivations to the Human-Data Interaction model~\cite{hdi-ssrn}, we each need a point-of-presence in this data ecology, providing:

\begin{itemize}
\item {\bf Legibility}. The ability to collate, inspect and reflect on ``our'' data, so we can understand what data is being collected and how it is being processed.
\item {\bf Agency}. The ability to control and manage ``our'' data and access to it, so we can have the capacity to act effectively in these systems as we see fit.
\item {\bf Negotiability}. The ability to continually navigate our way through the ways that data is social in construction and use (consider how little of ``your'' data concerns \emph{only} you and no-one else).
\end{itemize}

However, in providing these mechanisms it is critical to realise the enormous heterogeneity in users and attitudes: while some will want detailed engagement with many forms and uses of their data, others will not care. Thus the last point, negotiability, is key: ways in which we can set and manage policies expressing what we want to happen and providing a way to negotiate with other involved subjects, that we can interact with and mutate over time is at least as important as the more obvious requirements for mechanisms allowing us to see what is known and used about us, and to control the collection and use of such data. 

%% file: databox.tex
\section{What is a Databox?}
\label{s:databox}


So just what is a \databox? That is, what are the features and capabilities that one should provide? We divide our answer into four parts: it must be a \emph{trusted platform} providing facilities for \emph{data management} of data at rest for the data subjects as well as \emph{controlled access} for other parties wishing to use their data, and \emph{supporting incentives} for all parties.

\subsection{Trusted Platform}

Your \databox sits at the heart of your online presence. It captures, indexes, stores and manages data about you and data generated by you. To do so, you will have to trust it a great deal. As well as manually adding data and indexes into your \databox, data can be inferred from a variety of sources such as installed apps, browsing habits and online behaviour, in a \emph{privacy-conscious} manner. This potentially makes it a much more knowledgeable and intrusive system (albeit more useful), when compared to traditional data silos such as Amazon, Spotify and Google, and thus imposes requirements to protect users' privacy~\cite{mobiad}.

Trust in the platform also requires reliable behaviour as a piece of infrastructure. That is, a \databox must be consistently available so that it can usefully help the user manage their online interactions. At the same time it must provide straightforward means for the user to intervene in the data collection and sharing operations it is carrying out, to prevent breach in cases where automatic actions derived from configuration and policy have unforeseen consequences.

Finally, all of these actions and behaviours must be supported by pervasive logging, with associated tools, so that users and (potentially) third-party auditors can build trust that the system is operating as expected and, should something unforeseen happen, the results can at least be tracked.

\subsection{Controlled Access}

The purpose of a \databox is \emph{not} simply to gather all your personal data into one place, but to enable controlled access to that data. By this we mean that it must be selectively query-able: users should have fine-grained control over what data are made available to third parties. More complex possibilities include supporting privacy-preserving data analytics techniques such as differential privacy~\cite{diffpriv} and homomorphic encryption~\cite{Naehrig:2011:HEP:2046660.2046682}.

One important factor, often fleetingly considered (if at all!) by current systems, is the need to control the access period on a per-case basis. Specifically, the need to revoke previously granted access. In a system where access is granted to process data locally but not to take copies of data, this is relatively straightforward~\cite{dataware}; but in a system where data is, by default, copied out to the third-party, cooperation on their part is required to implement something like a \emph{time-to-live} function for data. A challenge in this space is the difficulty in measuring the impact of release of any given datum as it will be difficult at best for the \databox to maintain or otherwise obtain the existing and future information-states of all potential third parties that might access the newly released datum.

\subsection{Data Management}

As well as collating your data and providing means for granting controlled access to them, a \databox must provide means for users to interact with and reflect upon the data it contains. This will enable users to make more informed decisions about the behaviours they implement, whether directly themselves or indirectly by passing off control to others.

As part of these interactions, and to support trust in the platform, users must be able to edit and delete data from their \databox as a way to handle the inevitable cases where bad data is uncovered or discovered to have been inferred and distributed. Similarly, it may be appropriate for some data and desirable for some users to have the \databox \emph{not} exhibit the usual digital tendency of perfect record. Means to enable the \databox automatically to forget data that are no longer relevant or have become untrue may act as another factor increasing trust in the platform by users~\cite{mayer2009delete}. Even if data has previously been used, it may still need to be `put beyond use' by users who wish to redact it for the future~\cite{898852}. Such local and global concepts as the \emph{Right to be Forgotten} require adherence to agreed protocols. and other forms of cooperation, by third-party services and data aggregators.

\subsection{Supporting Incentives}

Development of innovative uses of personal data requires incentives. A consequence of the controlled access envisioned above is that users may choose to deny third-party services (e.g.,~advertisers or cloud service providers) access to their data. In the simplest case this might lead to those users simply no longer being able to make use of those services. However, a more acceptable and scalable option would rather be to provide means for those services to charge the user in other ways: those who wish to pay through access to their data may do so, while those who do not may pay through more traditional monetary means. That is, the \databox must be able to `talk money', enabling users to trace payments alongside data flow to and from different third-party services, available via some form of \emph{app store}.

The \databox could also act as an \emph{exposure reduction} mechanism for commercial organisations which may no longer intend to hold and control a range of private data directly (e.g., health records), and rather let the data subject take control of their sensitive information. The commercial organisation could still access and query the data as previously described. This is particularly relevant for international organisations that otherwise have to be aware of a plethora of legal frameworks. An analogy might be the way online stores use third-party payment services such as PayPal or Google Wallet to avoid the overhead of Payment Card Infrastructure compliance\footnote{Payment Card Infrastructure standards, \url{https://www.pcisecuritystandards.org/}} for processing credit card fees.

%% file: contents.tex
\section{What's in the Databox?}
\label{s:contents}

As soon as one begins to examine the requirements for a \databox, one thing becomes very clear: \emph{data} is a dangerous word. In particular, personal data is so complex and rich that treating it homogeneously is almost always a mistake. Various of the authors have attempted at various times to collate their digital footprints, and it proves a remarkably complex task. Subsequently deciding which devices should be able to share in and access the digital footprint, even before considering sharing with other people, makes it even harder. Issues such as mixed data formats (potentially proprietary), high variability in datum sizes, the multiplicity of standards for authentication to different systems to retrieve data (even within a single sector, e.g.,~banking), lack of standard data processing pipelines and tools, and myriad other reasons make this job infinitely fiddly -- none of these problems are inherently difficult but actually assembling and then maintaining the tools and data together inevitably takes considerable time and effort.

By way of example, one of the authors recently went through this exercise again. The (partial) footprint that resulted is over 55GB in size, with data from different sources spanning times from yesterday to over 10 years ago. Data types recovered include:

\begin{itemize}
\item
  {\bf Communications}. Email, Instant Messaging (over 6 services, some of which accounts have been idle for several years), phone call records, SMS exchanges.
\item
  {\bf Financial}. Bank statements (both personal and joint accounts), credit card statements, housing contracts/mortgage details.
\item
  {\bf Family}. Photographs (some of which contain family, some contain location metadata), trips, household energy consumption, shared calendars, children's health records.
\item
  {\bf Individual}. Personal location traces, personal calendars, address books, sleep tracking data.
\item
  {\bf Online Social Networks}. The usual candidates (Twitter, Facebook, Google+) as well as those no longer in existence (Orkut).
\end{itemize}

This data is initially collated on a reasonably powerful computer, with ample storage, CPU, and memory. But access to this data is desirable from a range of other devices including remote machines, tablets, and smartphones. While sync protocols such as BitTorrent Sync\footnote{\url{http://www.getsync.com/}} are approaching an adequately straightforward way to collate a lot of this data, it soon becomes clear that it is not so straightforward in practice. Factors such as different device capacities and capabilities mean that simply copying all this data to all devices is not a viable policy even when simply managing one's own data among one's own devices. However this limitation would also be a security and privacy minimising option, as one would at most have one or two strongly trusted -- i.e.,~utilising trusted hardware under the user's full control -- devices with access to the complete index of the data, with all other devices sending only limited queries to the trusted sources.

%% file: where.tex
\section{Where is my Databox?}
\label{s:where}


Having laid out motivation for \databox for all, and briefly explored some of the requirements and practicalities, it is natural to next ask: so where is mine? There have been several attempts to build systems that provide some or all of these features, but none have really been successful. We believe that this is because there are fundamental barriers, technical and social, that have yet to be successfully addressed.

{\bf Availability}.
  If the \databox is going to take such a central place in our online lives, then we cannot afford for it to become unreachable. This means that, as a network connected device, my \databox must be (securely) accessible no matter where I am; and it must also be itself reliable and robust against loss of power, connectivity, etc.

  The limitations imposed by extensive use of firewalls, NATs and other middlebox features in the current Internet have pushed past approaches to focus on use of the cloud to ensure connectivity under the assumption that connectivity to Internet-hosted servers is more widely available and reliable than connectivity between devices at the edge of the network. Pushing all connectivity to be via the cloud mitigates many of the problems introduced by middleboxes but brings a host of other issues, notably trust and cost.

{\bf Trust}.
  This is a multi-faceted aspect of a system. Two key aspects stand out: \one~the need to trust that the \databox will protect the user against breach of data due to, e.g.,~repeated queries or inference across different datasets; and \two~the need to trust that the software running on the \databox is trustworthy and not acting maliciously -- open source and virtualization or other sandboxing technologies, seem likely to have a key role to play here.

  More broadly, there is a need for uptake of such a service to begin somewhere -- how are the early adopters to be encouraged to use the facility, and once they start, how is this trust in the facility to be represented and propagated to others. Early experiences with both past attempts at personal data management systems as well as others such as online social networks, also suggests that trust in these systems is more complex than simply providing perfect recall: while we might be happy for our \databox to record everything perfectly in private, we might expect it to ``forget'' data over time, at least as far as others are concerned.

{\bf Complexity}.
  Existing systems intended to help users manage their personal data have found it difficult to control the associated complexity. User preferences in this space are inherently complex: socially derived and context dependent. They need to be expressed in machine-readable form so that software can assist us in this management, while also capturing the very broad range of intents and requirements. Two particular examples capture some of the inherent difficulties here. First, a three-year German study ending in 2012\footnote{\url{http://www.nytimes.com/2014/10/05/sunday-review/we-want-privacy-but-cant-stop-sharing.html}} showed that the more people disclosed about themselves on social media, the more privacy they said they desired. Sabine Trepte (the lead author of the study) observed that the paradox indicated dissatisfaction from the participants with what they got in return for giving away so much about themselves. And yet, she added, ``{\it they continued to participate because they were afraid of being left out or judged by others as unplugged and unengaged losers}''.

Second, many data are inherently, rather than explicitly, shared in that they implicate more than one individual. Common examples include domestic energy consumption data, and use of cloud email services such as Gmail: even if a user opts out by choosing not to use Gmail, there is a high chance\footnote{A 2014 analysis showed that ``51\% of the emails [that the author] replied to arrived from Google.'', making hiding information from Google impractical for the lay person, \url{http://mako.cc/copyrighteous/google-has-most-of-my-email-because-it-has-all-of-yours}} that the recipient of their email \emph{is} using Gmail and so the sender cannot prevent Google discovering the contents of their message. It is thus not always clear who \emph{owns} which piece of data or has the right to grant permissions to a shared data item.

{\bf Usability}.
  Related to the issue of complexity is that of usability, one area in which the centralised platform providers have excelled. The complexity that is inherent to the systems being created needs to be made legible, empowering end-users to understand the choices they have available and the consequences of their actions. A successful \databox will need consistent user interaction models and will enable developers of \databox applications to make use of these models. One area of inspiration here might be the work done in the Homework project which prototyped and, through a number of deployments of several months each, studied use of several novel task-specific interfaces assisting users in the complex business of understanding and managing their home broadband networks~\cite{homework}. It is worth noting that the high penetration of broadband and even larger adoption of smartphones, individuals today are substantially more sophisticated than the \emph{naive} users of the past, and capable of embracing the data management capabilities of a \databox. Moreover, due to large coverage of privacy and personal data issues in the media, users may be seeking out solutions like a \databox but without sacrificing the user-experiences they have become familiar with.

{\bf Cost}.
  As always, with a new facility such as a \databox, there are a range of incentives that need to be aligned for success. Operational costs of running a \databox have to be acceptable to users. Coupled with this, the costs of third-parties accessing the system, and potentially having to recompense for access to data that previously they would've simply gathered, will have to be recouped. It remains to be seen how this can be done in practice: Are users willing and able to pay in practice? What will be the response of users when offered pay-for versions of previously free-to-use services? There is some evidence that at least some users will be willing to make this trade-off, but the same studies also show that the situation is complex~\cite{skatova13:perceiv}.

%% file: conclusion.tex
\section{When can I have my Databox?}
\label{s:conclusion}



Having made a case for each of us to have a \databox, the only remaining question is when can we have it? In a market-based economy, as ever, this requires the right combination of sufficiently high demand and sufficiently low cost that the need can be met. We are pursuing reduction in cost through development of associated technologies, including Nymote\footnote{\url{http://nymote.org}} and its constituent components including Mirage~\cite{mirage-asplos}, Irmin~\cite{irmin-oud} and Signpost~\cite{signpost-foci}. In addition, we are developing methodologies for indexing and tracking the personal data held about us by third parties.

However, the problem of demand highlights several unresolved challenges in this space: if, by and large, people do not see the need for technologies like this unless and until they suffer some kind of harm from a data breach, it may remain difficult to reach sufficiently high levels of demand; though this may change as public education programs proceed to teach people about the potentials and challenges of personal data. It is even possible that governments will feel compelled to regulate to protect their citizens even before there is clear popular demand: some have posited, for example, a need to change consent models from current practices of obtaining ``informed consent'' (be it never so informed) to something more akin to ``consumer protection'' (you do not give informed consent to buy food from a supermarket -- you assume that the food on sale is generally fit for consumption)~\cite{luger2013informed}.

From a technology design point of view, the general approach proposed is that of ``privacy by design'' (PbD): it remains to be seen whether PbD can be successfully implemented in a space such as this where policy and technology need to co-evolve. Even then, there also needs to be an explicit involvement of the social aspects: it is unlikely that either state of everything-public or everything-hidden is desirable for society. Ultimately, the litmus test of success for personal data containers will be their wide-scale adoption and operation. In order to evaluate their effectiveness and the possibility of release or sale of personal data, there needs to be a method for determining the marginal rate of substitution\footnote{This measures the rate at which the consumer is just willing to substitute one good for another.} for personal data. The sale of personal data and the rich insights and analytics derived from it is considered the key utility in this ecosystem, and the individuals' preferences are the fundamental descriptors and success indicators. Perhaps availability of such rich and indexed data in one central aggregation point would enable one to build a digital image of myself from the outside world point of view, depending on what information is released to which external actors.

Even observing a number of individuals using such a tool in the wild will enable understanding of their real willingness to pay for services, or marginal willingness to pay for privacy. It has been argued that privacy is negotiated through collective dynamics, and hence society reacts to the systems that are developed and released~\cite{Gurses:2014:YEP:2632661.2633029}. This calls for trial deployments and in-the-wild studies of personal data containers in partnership with individuals, in addition to successful negotiation with consumer rights groups, privacy advocates, the advertising industry, lawmakers, and regulators. Considering the churn experienced in the personal data startup space, with a number of new but typically short-lived entrants and offering, it seems that few truly viable business models have yet been discovered in this space. Our belief is that the power of personal data can only be realised when proper consideration is given to its social character, and it can be legibly and negotiably combined with data from external sources. In this case, we might anticipate many potential business models~\cite{hat}.

%% file: percon14.bbl
\begin{thebibliography}{10}

\bibitem{Balebako12measuringthe}
Balebako, R., Leon, P.~G., Shay, R., Ur, B., Wang, Y., and Cranor, L.~F.
\newblock Measuring the effectiveness of privacy tools for limiting behavioral
  advertising.
\newblock In {\em In Web 2.0 Workshop on Security and Privacy} (2012).

\bibitem{898852}
Brown, I., and Laurie, B.
\newblock Security against compelled disclosure.
\newblock In {\em Computer Security Applications, 2000. ACSAC '00. 16th Annual
  Conference} (Dec 2000), 2--10.

\bibitem{10.1371/journal.pone.0098790}
de~Montjoye, Y.-A., Shmueli, E., Wang, S.~S., and Pentland, A.~S.
\newblock openpds: Protecting the privacy of metadata through safeanswers.
\newblock {\em PLoS ONE 9}, 7 (07 2014).

\bibitem{dourish-context}
Dourish, P.
\newblock What we talk about when we talk about context.
\newblock {\em Personal Ubiquitous Comput. 8}, 1 (Feb. 2004), 19--30.

\bibitem{diffpriv}
Dwork, C.
\newblock Differential privacy.
\newblock In {\em Automata, Languages and Programming}, M.~Bugliesi,
  B.~Preneel, V.~Sassone, and I.~Wegener, Eds., vol.~4052 of {\em LNCS}.
  Springer Berlin / Heidelberg, 2006, 1--12.

\bibitem{DBLP:journals/corr/FalahrastegarHUM14}
Falahrastegar, M., Haddadi, H., Uhlig, S., and Mortier, R.
\newblock Anatomy of the third-party web tracking ecosystem.
\newblock {\em CoRR abs/1409.1066\/} (2014).

\bibitem{irmin-oud}
Gazagnaire, T., Chaudhry, A., Crowcroft, J., Madhavapeddy, A., Mortier, R.,
  Scott, D., Sheets, D., and Tsipenyuk, G.
\newblock Irmin: a branch-consistent distributed library database.
\newblock In {\em Proceedings ICFP OCaml User and Developer Workshop} (Sept.~5
  2014).

\bibitem{guha.hotnets09}
Guha, S., Reznichenko, A., Tang, K., Haddadi, H., and Francis, P.
\newblock Serving ads from localhost for performance, privacy, and profit.
\newblock In {\em {ACM Workshop on Hot Topics in Networks}} (2009).

\bibitem{Gurses:2014:YEP:2632661.2633029}
G\"{u}rses, S.
\newblock Can you engineer privacy?
\newblock {\em Commun. ACM 57}, 8 (Aug. 2014), 20--23.

\bibitem{mobiad}
Haddadi, H., Hui, P., and Brown, I.
\newblock Mobiad: private and scalable mobile advertising.
\newblock ACM MobiArch (2010).

\bibitem{privalytics}
Haddadi, H., Mortier, R., Hand, S., Brown, I., Yoneki, E., McAuley, D., and
  Crowcroft, J.
\newblock Privacy analytics.
\newblock {\em ACM Computer Communication Review\/} (April 2012).

\bibitem{dontkill}
Leontiadis, I., Efstratiou, C., Picone, M., and Mascolo, C.
\newblock Don't kill my ads!: balancing privacy in an ad-supported mobile
  application market.
\newblock ACM HotMobile (2012).

\bibitem{luger2013informed}
Luger, E., and Rodden, T.
\newblock An informed view on consent for ubicomp.
\newblock In {\em Proceedings of the 2013 ACM international joint conference on
  Pervasive and ubiquitous computing}, ACM (2013), 529--538.

\bibitem{mirage-asplos}
Madhavapeddy, A., Mortier, R., Rotsos, C., Scott, D., Singh, B., Gazagnaire,
  T., Smith, S., Hand, S., and Crowcroft, J.
\newblock Unikernels: Library operating systems for the cloud.
\newblock In {\em Proceedings 18th International Conference on Architectural
  Support for Programming Languages and Operating Systems (ASPLOS)} (Houston,
  TX, USA, Mar.~16--20 2013).

\bibitem{mayer2009delete}
Mayer-Schonberger, V.
\newblock {\em Delete: The Virtue of Forgetting in the Digital Age}.
\newblock Princeton University Press, 2009.

\bibitem{dataware}
McAuley, D., Mortier, R., and Goulding, J.
\newblock {The Dataware Manifesto}.
\newblock In {\em Proceedings of the 3rd IEEE International Conference on
  Communication Systems and Networks (COMSNETS)} (Bangalore, India, January
  2011).
\newblock Invited paper.

\bibitem{mortier2010personal}
Mortier, R., Greenhalgh, C., McAuley, D., Spence, A., Madhavapeddy, A.,
  Crowcroft, J., and Hand, S.
\newblock The personal container, or your life in bits.
\newblock {\em Proceedings of Digital Futures\/} (2010).

\bibitem{hdi-ssrn}
Mortier, R., Haddadi, H., Henderson, T., McAuley, D., and Crowcroft, J.
\newblock Human-data interaction: The human face of the data-driven society.
\newblock {\em SSRN\/} (Oct.~1 2014).
\newblock \url{http://dx.doi.org/10.2139/ssrn.2508051}.

\bibitem{homework}
Mortier, R., Rodden, T., Tolmie, P., Lodge, T., Spencer, R., Crabtree, A.,
  Sventek, J., and Koliousis, A.
\newblock Homework: Putting interaction into the infrastructure.
\newblock In {\em Proceedings of the 25th Annual ACM Symposium on User
  Interface Software and Technology}, UIST '12, ACM (New York, NY, USA, 2012),
  197--206.

\bibitem{Mun:2010:PDV:1921168.1921191}
Mun, M., Hao, S., Mishra, N., Shilton, K., Burke, J., Estrin, D., Hansen, M.,
  and Govindan, R.
\newblock Personal data vaults: A locus of control for personal data streams.
\newblock In {\em Proceedings of the 6th International COnference}, Co-NEXT
  '10, ACM (New York, NY, USA, 2010), 17:1--17:12.

\bibitem{Naehrig:2011:HEP:2046660.2046682}
Naehrig, M., Lauter, K., and Vaikuntanathan, V.
\newblock Can homomorphic encryption be practical?
\newblock In {\em Proceedings of the 3rd ACM Workshop on Cloud Computing
  Security Workshop}, CCSW '11, ACM (New York, NY, USA, 2011), 113--124.

\bibitem{hat}
Ng, I.~C.
\newblock {Engineering a Market for Personal Data: The Hub-of-all-Things (HAT),
  A Briefing Paper }.
\newblock {\em WMG Service Systems Research Group Working Paper Series\/}
  (2014).

\bibitem{signpost-foci}
Rotsos, C., Howard, H., Sheets, D., Mortier, R., Madhavapeddy, A., Chaudhry,
  A., and Crowcroft, J.
\newblock Lost in the edge: Finding your way with signposts.
\newblock In {\em Proceedings 3rd USENIX Workshop on Free and Open
  Communications on the Internet (FOCI)} (Washington D.C., USA, Aug.~13 2013).

\bibitem{skatova13:perceiv}
Skatova, A., Johal, J., Houghton, R., Mortier, R., Bhandari, N., Lodge, T.,
  Wagner, C., Goulding, J., Crowcroft, J., and Madhavapeddy, A.
\newblock Perceived risks of personal data sharing.
\newblock In {\em Proceedings Digital Economy 2013: Open Digital} (Nov. 2013).

\bibitem{Vulnerable2014}
Urban, J.~M., and Hoofnagle, C.~J.
\newblock The privacy pragmatic as privacy vulnerable.
\newblock In {\em Symposium on Usable Privacy and Security (SOUPS 2014)
  Workshop on Privacy Personas and Segmentation (PPS)} (July 2014).

\bibitem{narseoIMC}
Vallina-Rodriguez, N., Shah, J., Finamore, A., Grunenberger, Y., Papagiannaki,
  K., Haddadi, H., and Crowcroft, J.
\newblock Commercial break: Characterizing mobile advertising.
\newblock In {\em ACM SIGCOMM Internet measurement conference} (2012).

\bibitem{westin-survey}
Westin, A.~F.
\newblock {\em E-commerce \& Privacy: What Net Users Want}.
\newblock Privacy \& American Business, Hackensack, NJ, 1998.

\end{thebibliography}
